\documentclass[traditabstract]{aa}
\usepackage[utf8]{inputenc}
\usepackage{amsmath}
\usepackage{amssymb}
\usepackage{microtype}
\usepackage{graphicx}
\usepackage{txfonts}
\usepackage{color}
\usepackage{multirow}
\usepackage{xspace}
\usepackage{tabularx}
\usepackage{rotating}
\usepackage{longtable}
\usepackage{ulem}
\usepackage{verbatim}
\usepackage{array}
\newcolumntype{L}[1]{>{\raggedright\let\newline\\\arraybackslash\hspace{0pt}}m{#1}}

\bibpunct{(}{)}{;}{a}{}{,} 

\usepackage{natbib}
 \bibpunct{(}{)}{;}{a}{,}{,} 

\begin{document}

\title{Distances to Accreting X-ray pulsars: impact of the Gaia DR2.}
\author{S.~Treuz\inst{1}, V.\,Doroshenko\inst{1}, A.\,Santangelo\inst{1}, R.\,Staubert\inst{1}}  

\institute{$^1$Institut für Astronomie und Astrophysik, Universität Tübingen, Sand 1, 72076 Tübingen, Germany
}

\abstract{
A well constrained estimate of the distance remains one of the main factors to properly interpret
 the observations of accreting X-ray pulsars. Since these
objects are typically well studied, multiple distance estimates obtained with
different methods are available for many members of the class. Here we summarize all
 distance estimates published in the literature for a sample of Galactic X-ray
pulsars, and compare them with distance measurements
obtained by the Gaia mission. We conclude that the spread of distance values obtained
for individual objects by the different conventional methods is usually larger than one might expect
from the quoted individual uncertainties, such that Gaia values are in many cases a very useful additional information. 
For distances larger than 5 kpc, however, the uncertainties of all distance estimates (including those of Gaia) remain comparatively large, so conventional methods will likely retain their importance. 
We provide, therefore, an aposteriori estimate of the systematic uncertainty for each method based on
the comparison with the more accurate Gaia distance measurements.
}

\authorrunning{S. Treuz et al.}
\maketitle

\section{Introduction}
The emission of X-ray pulsars is powered by accretion of matter captured
from a non-degenerate, typically massive companion star. Their observed properties
are largely defined by the accretion rate, which is directly related to
the observed X-ray luminosity. Similarly to many other astronomical objects,
the knowledge of distance is, therefore, essential for properly interpreting their observed
phenomenology \citep{1975AandA....42..311B,2009arXiv0912.0252C}.

On the other hand, distance estimation is rather challenging for accreting X-ray pulsars,
and in most cases is deduced for the identified optical counterpart. Given that
most of the known objects are fairly bright and remote, it usually was not
possible to measure the parallax of the optical star before Gaia, so
alternative methods had to be used. As briefly summarized below, these are in
most cases model dependent, and as such might be subject to systematic
uncertainties potentially affecting many conclusions regarding the physical properties
 of X-ray pulsars.
On the other hand, the strong interest of the community in accurate distances to X-ray pulsars
has triggered multiple dedicated investigations. For many objects several
distance estimates with independent methods are available.

In this paper we compare the available distance estimates with the
direct parallax measurements obtained for the first time by the Gaia mission \citep{2016AandA...595A...1G} and analysed by  \cite{2018arXiv180410121B}. The
 paper is organized as follows. In section \ref{sec:data_methods} we
summarize the methods for distance estimates used in the literature for X-ray
pulsars, and the results obtained for a sample of objects for which such
estimates are available. In section \ref{sec:conclusions} we then compare these
with the results of Gaia Data Release 2 \citep{2018arXiv180409365G}, assess the accuracy and biases associated with
individual methods, and briefly discuss possible implications of the updated
distances on the physical properties of several accreting pulsars.

\section{Data and methods}
\label{sec:data_methods}
The main goal of this study is to compare the new distances provided by the Gaia
mission with already available estimates, specifically for the case of X-ray
pulsars. To this aim we considered only objects for which the distance has already been
estimated, and Gaia parallaxes are available. As a starting step, we
conducted a literature search for all X-ray pulsars listed in the catalogue made
available by Mauro
Orlandini\footnote{\url{http://www.iasfbo.inaf.it/~mauro/pulsar_list.html}}, 
which have been reported to exhibit cyclotron resonance scattering features. We
compiled a list of objects with individual distance etimates. We
omitted cases for which the only distance estimate available was based on the
location in the Galactic plane and possible association with one or several
Galactic arms. The final list of sources, including the individual distance
estimates, corresponding references and Gaia counterparts is presented in
Table~\ref{tab:src}. The table also contains information on how the distance
was estimated, denoted with one of the following codes:

\begin{itemize}
\item \textbf{Spectro-Photometry}
In all methods of the following three groups, the spectral class and effective temperature of the companion is determined using optical spectroscopy.
The distance is then deduced based on the photometry. 
\begin{itemize}
    \item \textbf{SPH} - \textbf{S}pectro-\textbf{Ph}otometry. The reddening of the optical companion is determined e.g., by inferring true colors from the spectral type and comparing them with photometric data \citep{2001AandA...371.1018I} or measuring the EW of interstellar absorption lines \citep{2005AandA...440.1079R}. A reddening law is assumed (usually 2.75<R<3.5) to determine the total extinction. Single or multiple band photometric data are then used to determine the distance. Uncertainties for this method
    arise from the assumed extinction law, the intrinsic spectrum of the star, the method used to determine reddening, and the possibility
    of heating by the compact object, which could affect the identification of the spectral class.
    \item \textbf{SED}-Modeling of the \textbf{s}pectral \textbf{e}nergy \textbf{d}istribution. In this method multiband photometric data are fitted by a blackbody using a temperature inferred by the spectral class \citep[see][]{2013ApJ...764..185C}.
    \item \textbf{ATM}-\textbf{Atm}ospheric model. A Kurucz model stellar atmosphere is fitted to the spectrum of the optical counterpart incorporating the full spectrophotographic data into the distance estimation \citep[see][]{2014ApJ...793...79L}.
\end{itemize}

 \item \textbf{NH}-Absorbtion and exctinction $\textbf{N}_{\textbf{H}}$. Assuming that the distribution of the absorbing material
    in the Galaxy is known, one can estimate the distance to the source based on the observed absorption in 
    X-rays or on optical exctinction \citep{1981MNRAS.195P..67M}.  Uncertanties arise from the poorly known dust distribution and composition, 
    from the extinction law, and the possibility of local absorption in the binary system.

\item \textbf{DH}-X-ray \textbf{d}ust scattering \textbf{h}alos. Interstellar dust scatters 
    X-rays which leads to the appearance of a ring-like halo around bright X-ray sources. Assuming that the angular dependence
    of the scattering cross-section and dust distribution along the line of sight are known, it is possible to deduce
    the distance to the source based on the observed shape of the halo \citep{2004ApJ...610..956C,1973AandA....25..445T}. Uncertainties arise from the assumed dust composition (which influences the scattering cross-sections), and distribution along the line of sight.

    \item \textbf{SU}-Spin evolution of the pulsars. The \textbf{s}pin-\textbf{u}p can be related to
    the X-ray luminosity yielding constraints on the distance as shown in
    \cite{Lipunov,Scott97,2017arXiv171010912D}. Distance estimates are affected by the uncertainty in the physical mechanisms
    driving the spin evolution of accreting neutron stars, and particularly spin-down torque.
    Both spin-up and spin-down torques also depend on parameters of the neutron star such
    as moment of inertia, magnetic moment and field configuration, which thus represent an additional source of uncertainty.

    \item \textbf{PX}-\textbf{P}aralla\textbf{x}. Parallaxes are the most direct way of distance determination in astronomy. Unfortunately
    X-ray pulsars are typically relatively far away, so before Gaia it was only possible to measure parallaxes
    for a few selected systems \citep[mostly with Hipparcos,][]{1998AandA...330..201C}.

    \item \textbf{CAII}-\textbf{Ca II} interstellar absorption line. The ratio of the
    equivalent widths of the Ca II K and H lines can be used to determine the
    distance \citep{2009AandA...507..833M}. This depends on the calibration
    and assumed model.

     \item \textbf{SV}-\textbf{S}ystemic \textbf{v}elocity. The observed radial velocity of the companion,
    determined from optical spectroscopy, can be attributed to the Galactic motion
    of the star, which allows to estimate its distance providing the rotation
    curve of the galaxy is known. Uncertainties arise from the rotation
    curve and peculiar velocity of the star \citep{2003RMxAA..39...17K}.

    \item \textbf{LC}-\textbf{L}ight \textbf{c}urve. \cite{1982APJ...262..253M} used the distance as
    a parameter to fit a multi-band light curve model describing additional exctinction
    by the thick outer rim of the accretion disc to the observational data.
    This and similar estimates are higly model dependent.
    
    \item \textbf{EB}-\textbf{E}clipsing \textbf{b}inaries. Similar to the method described above, but more reliable
    as the dimensions of the companion star can be constrained from
    spectroscopy and light curve analysis. With this method the parameters of the
    eclipsing binary can be determined, including the semimajor axis, masses of
    the companion, etc. The derived temperature and geometrical size of the
    donor star can then be used to infer the distance 
    \citep{1997MNRAS.288...43R}. 
    
    \item \textbf{RA}-\textbf{R}un\textbf{a}way star. Assuming an origin in a nearby cluster, the
    proper motion and age estimates can be used to determine the distance to the
    cluster, the distance of which can be determined by other means. This method
    can only be used in special cases as in \cite{2001AandA...370..170A}.

    \item \textbf{ACH}-\textbf{A}ccretion \textbf{c}olumn \textbf{h}eight. \cite{1998AdSpR..22..987M} have used
    the distance to cyclotron line sources to calculate the height of the emitting region.
    In reverse, by assuming a model for the dependence of the cyclotron energy,
    they could infer the distance. Given the poor understanding of the cyclotron line
    origin \citep[see e.g. discussion in][]{Poutanen13}, and the not yet fully understood physics of the
    accretion column \citep{Mushtukov15}, such estimates are strongly model
    dependent.

\end{itemize}

\section{Discussion and conclusions} \label{sec:conclusions} 

To compare the distance estimates of conventional methods to the new Gaia data, the 
measured parallaxes have to be converted to a distance estimate. As discussed in detail by
\cite{2015PASP..127..994B} and \cite{2018arXiv180409376L} , inversion of the parallax to distance is
only feasible if the parallax is well constrained. Even then, to obtain a reliable estimate for the resulting
distance a more sophisticated analysis, possibly using additional prior knowledge regarding
the considered objects, is required to account for systematic effects. 

\begin{figure}
\resizebox{\hsize}{!}{\includegraphics{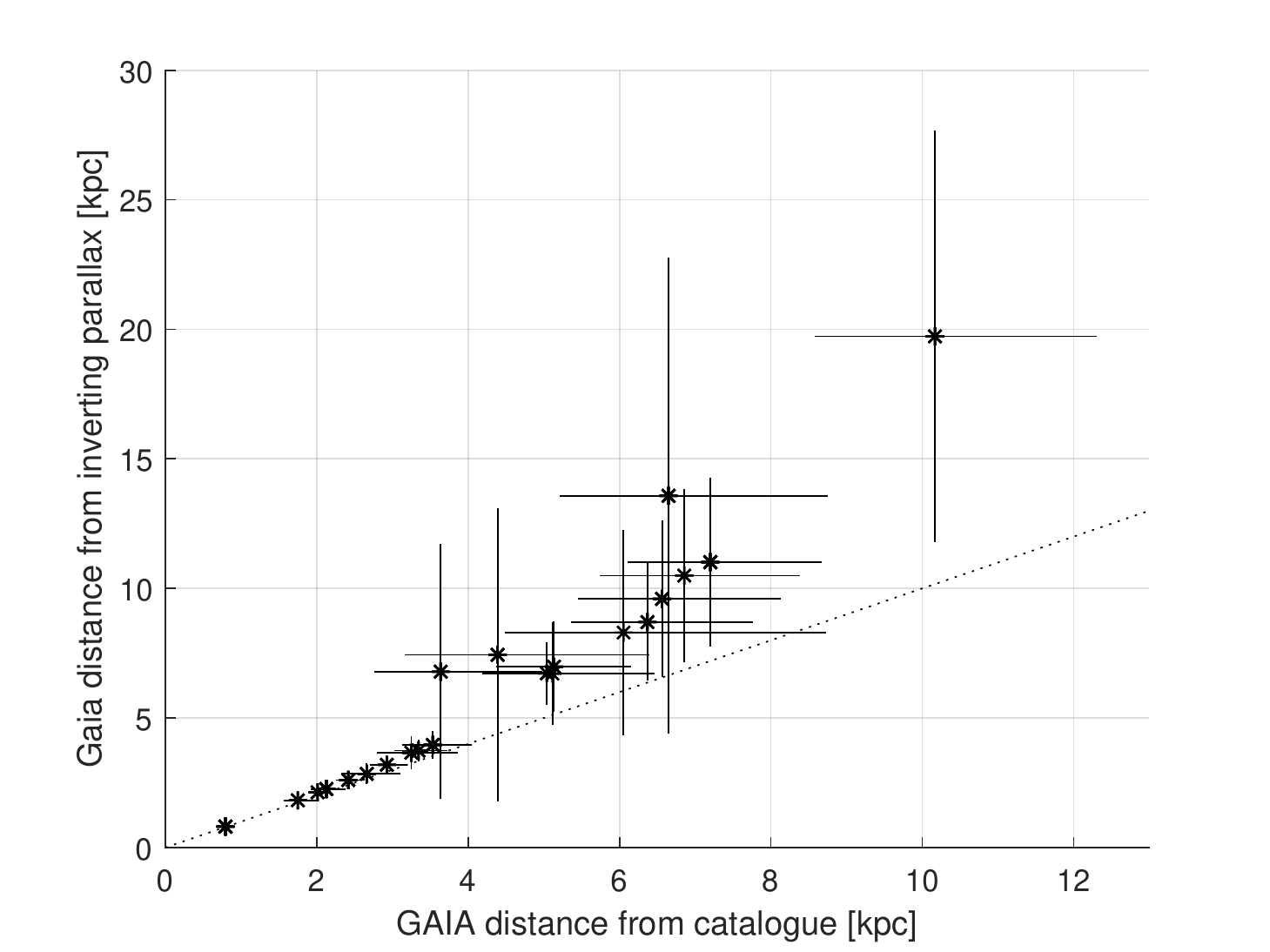}}
\caption{Comparison of distance estimates from inverting the Gaia parallax and the catalogue values \citep{2018arXiv180410121B} for our sample. The dotted line corresponds to $d_{px}=d_{BJ}$}
\label{fig:gaiavscat}
\end{figure}

For instance, \cite{2018arXiv180410121B} used a Bayesian probabilistic analysis to estimate distances for sources in Gaia DR2 considering a smoothed model of the
length scale distribution of the Galaxy as a prior. This approach yields more reliable estimates for distances and their uncertainties, compared to simple
parallax inversion. It is also claimed that meaningful distances can be obtained for measured parallaxes that are smaller than their uncertainties , and even negative parallaxes (which can be debated). The
distances for the sources in our sample obtained by \cite{2018arXiv180410121B} are listed in Table \ref{tab:src} together with an estimate by a simple inversion. The posterior
probability densities reported in this catalogue are unimodal for all sources in our sample. The estimate quoted in the table corresponds to the mode, the errors to a
1 $\sigma$ interval. Figure \ref{fig:gaiavscat} illustrates the difference in the two methods. Note that while the agreement is generally reasonable for distances $\lesssim$ 4 kpc, while for larger
values, the probabilistic approach yields somewhat smaller distances. This is due to the fact that corresponding parallaxes are smaller and thus poorly constrained
which leads to biased distance estimates. We use, therefore, only the distances reported by \cite{2018arXiv180410121B} from here on.

\begin{figure}
\resizebox{\hsize}{!}{\includegraphics{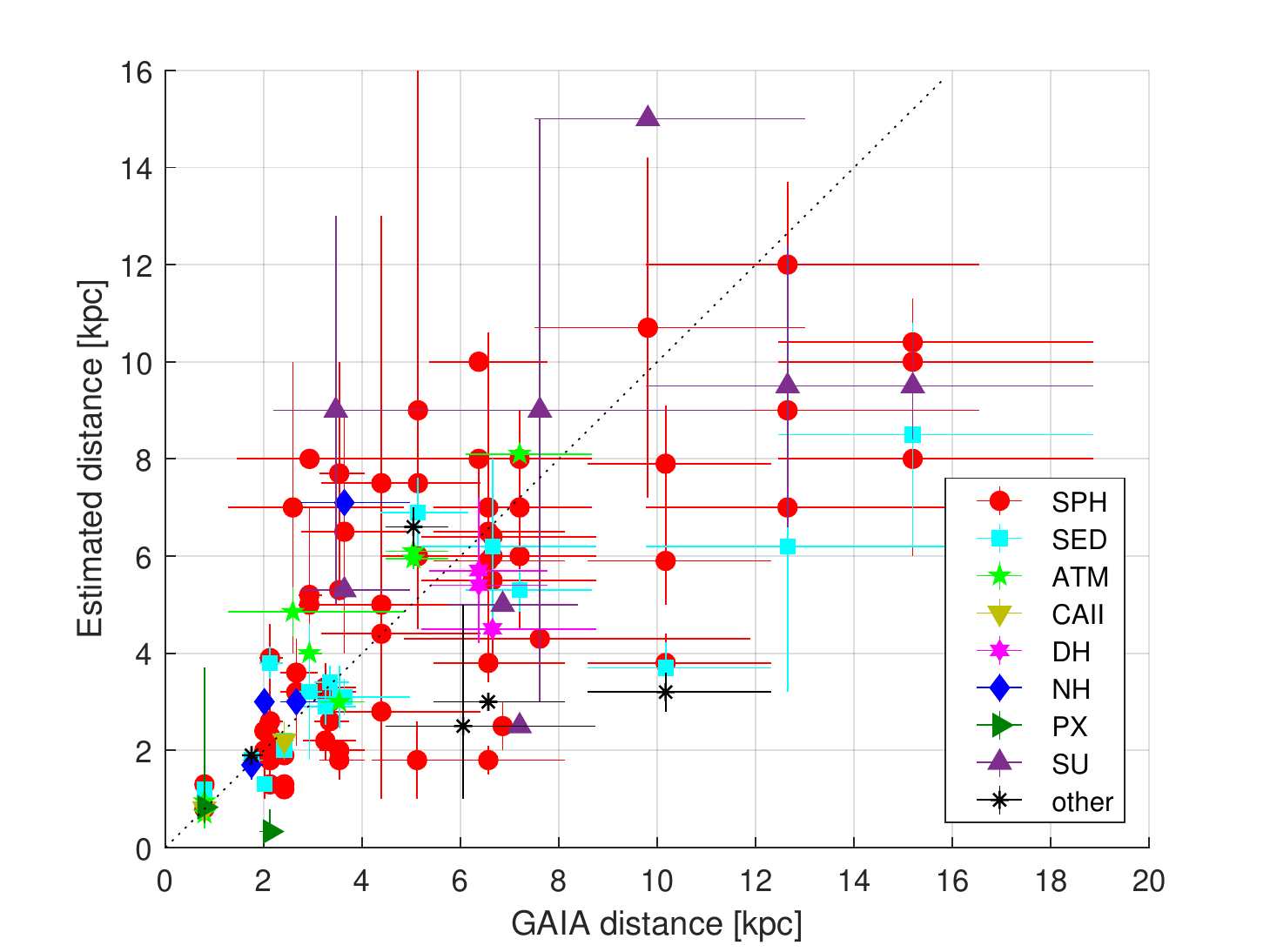}}
\caption{Comparison of distance estimates from the literature and GAIA distances \citep{2018arXiv180410121B} for our source sample. The symbols denote different methods used in the literature distance estimation described in section \ref{sec:data_methods}. The dotted line corresponds to $d_{est}=d_{BJ}$}
\label{fig:distance}
\end{figure}

\begin{figure}
\resizebox{\hsize}{!}{\includegraphics{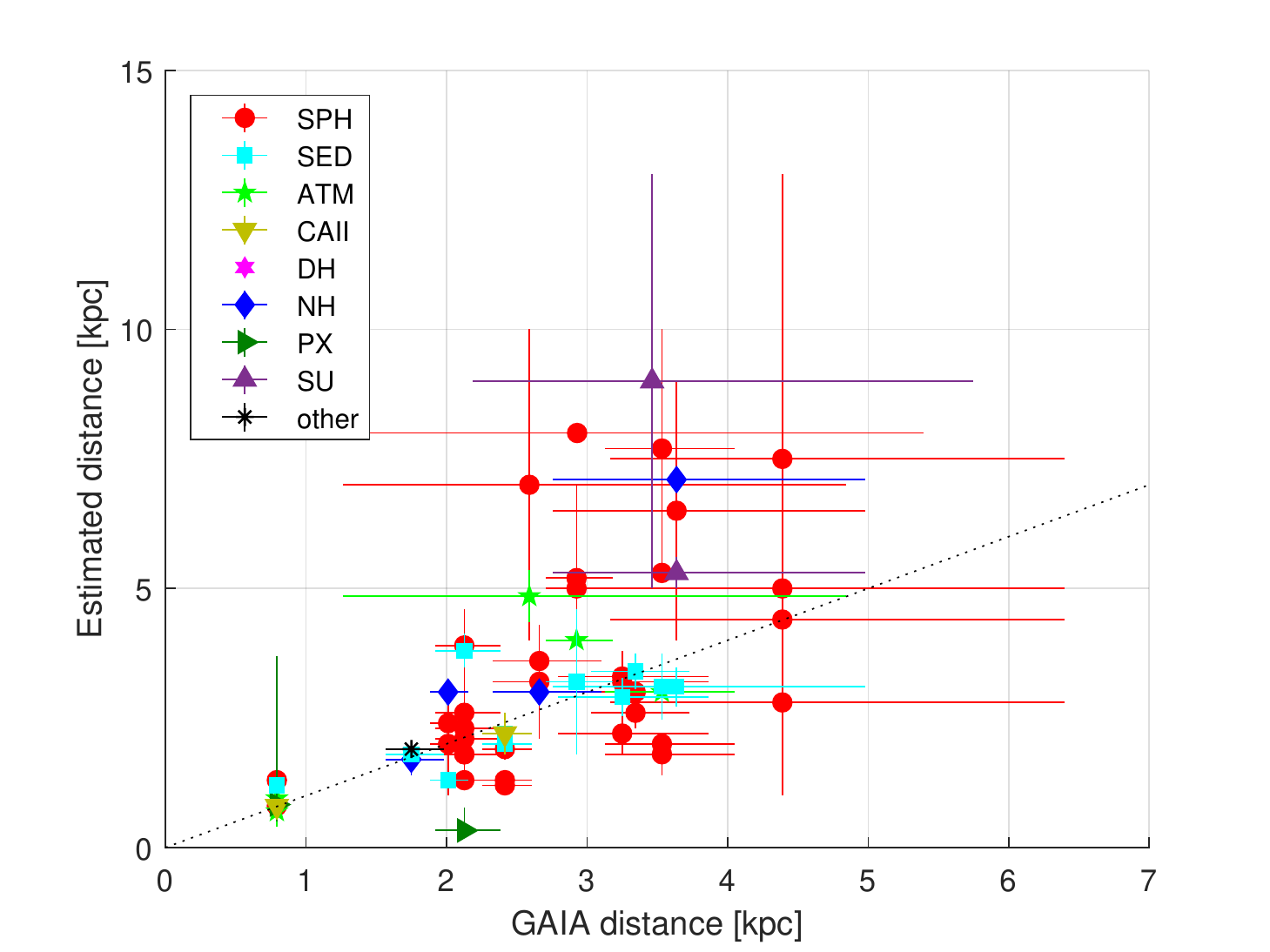}}
\caption{Detail of figure \ref {fig:distance} for GAIA distances \citep{2018arXiv180410121B} below 5 kpc. The symbols denote different methods used in the literature distance estimation described in section \ref{sec:data_methods}. The dotted line corresponds to $d_{est}=d_{BJ}$}
\label{fig:distance2}
\end{figure}

Figure \ref{fig:distance} shows the comparison between the distance estimates
gathered from the literature to the new GAIA DR2 distances. Different
symbols denote different estimation methods, while estimation methods with only
one example within our sample have been grouped into one symbol. All points are, however, listed in Table.~\ref{tab:src}. Figure
\ref{fig:distance2} shows a detail of figure \ref{fig:distance} for GAIA
parallax distances below 5 kpc. As it can be seen, distances estimated with the
aforementioned methods, while displaying a fairly large scatter, appear to
exhibit a slight systematic overestimation compared to Gaia results for Gaia distances $\lesssim$ 5 kpc. On the
contrary, for Gaia distances between 5 and 8 kpc, the distances reported by \cite{2018arXiv180410121B} appear to be
slightly overestimated. This is even stronger the case for the two sources with $d_{BJ} $> 12 kpc (KS 1947+300 and XTE J1946+274), for which the measured parallaxes are smaller than the corresponding uncertainty, or even negative, respectively. \\
To quantify these points, and to compare the performance of the different
methods, we calculated the relative distance difference as ($d_{est} -
d_{BJ}$)/$d_{BJ}$ for each source and averaged these for each
method. The results are presented in Table \ref{tab:stat}. The errors quoted in the literature for 
different estimates describe different confidence intervals or bounds, thus, no consistent weight for computing a weighted average could be chosen for all estimates.
Instead, an arithmetic mean was used. The errors correspond to the standard deviation of the distribution. 
Positive or negative values would point to a systematic overestimation or underestimation of any given method respectively.
However, the values for most methods are compatible with 0 within the standard deviation and those which are not (NH, DH) have
small sample size.  To quantify the distance dependent systematic seen in fig. \ref{fig:distance}, the average for all methods is also given in 
table \ref{tab:stat} for Gaia distance below and above 5kpc. The change from overestimation to underestimation is apparent. 
This systematic underestimation of distances by all conventional
methods for more distant sources is a bit alarming, and could point to a
sub-optimal choice of priors in the analysis by \cite{2018arXiv180410121B}.
This issue, however, requires a specific study using a larger and more diverse
sample of objects, and is out of scope of this work.

\begin{figure}
    \resizebox{\hsize}{!}{\includegraphics{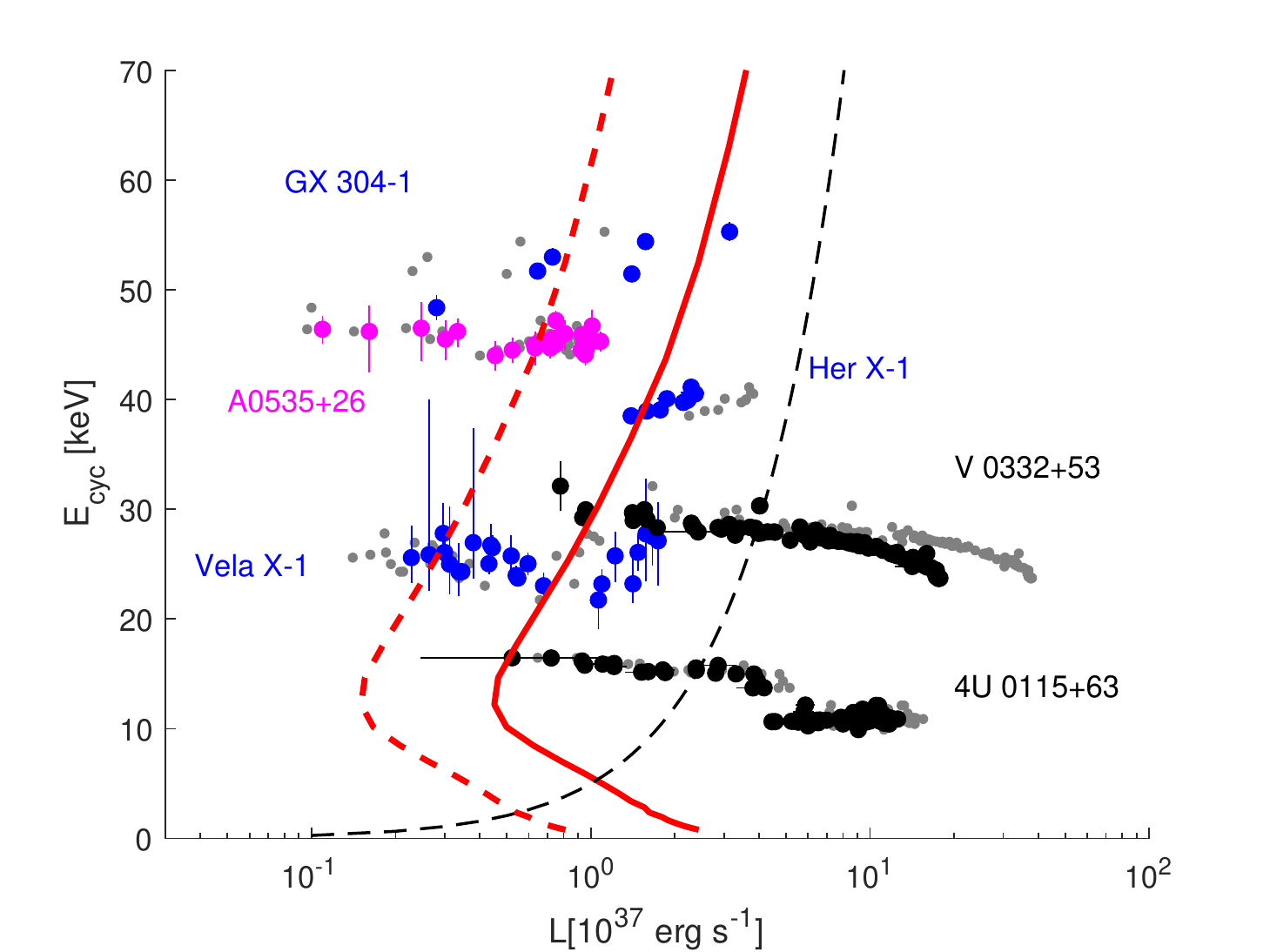}}
    \caption{Luminosity dependence of the fundamental cyclotron line energy. 
    Grey dots show previous work by \cite{2015MNRAS.447.1847M} using conventional
    distance estimates. The red dashed and solid curves show his suggested lower and upper bounds for the critical luminosity respectively. The filled circles show the data taken from \cite{2015MNRAS.447.1847M} corrected for the new GAIA DR2 parallax distances. The black dashed curve shows the predictions by \cite{2012AandA...544A.123B} }
    \label{fig:cyc}
\end{figure}

As mentioned before, the distance estimates obtained through conventional methods
display a relatively large spread for any given source, so using one of these distances 
to determine the properties of a star can potentially create large errors. To check 
the impact of the new Gaia distances, we investigated the changes in an effect relevant
to our sample of accreting X-ray pulsars exhibiting cyclotron resonance scattering features.   
Cyclotron lines arise due
to the resonant scattering of photons on electrons in strong magnetic field,
and thus trace the field strength in the line forming region. Dependence of the
observed line energy on luminosity is thus directly related to dimensions of
the emission regions, which can be used to probe the interaction of the
accretion flow with radiation and magnetic fields \citep{2015MNRAS.447.1847M}.
This dependence can also be used to better constrain the transitional
luminosity from the onset of the so-called accretion column \citep{Doroshenko17,Vybornov18}.
 A transition from positive to negative
correlation of the cyclotron line energy with flux is expected, and was actually observed
\citep{Doroshenko17} at certain critical luminosity, which can be used to
determine basic parameters of the neutron star provided that the line origin
and dependence of the accretion column parameters on the accretion rate are
understood. \cite{2015MNRAS.447.1847M} discussed this aspect and confronted
theoretical predictions with observations assuming the distances reported by
\cite{2012AandA...544A.123B} and \cite{2014ApJ...780..133F}.
 Figure \ref{fig:cyc}
shows for six accreting pulsars the luminosity dependence of the cyclotron line
centroids, obtained using the improved GAIA DR2 distance estimates as well as
in grey the old results from \cite{2015MNRAS.447.1847M}. The theoretical
predictions for the transitional luminosity, also shown in the figure,
correspond to different assumptions regarding the geometry of the column, and
scattering cross-sections. As it can be seen, differences are minor.\\
In conclusion, we compared distance estimates from the literature with new Gaia distances
and found a systematic overestimation at distances < 5kpc and underestimation at distances > 5kpc 
for all conventional methods which might point to a problem in the distance determination from the 
parallaxes. We compared the accuracy of different methods, but the large scatter dominates any systematics 
of individual methods.  Lastly, we investigated the implication of the new Gaia distances for previous work on 
the luminosity dependence of the cyclotron line. The differences turned out to be minor, as the relevant sources
were very well studied.  
We conclude, that conventional distance determination methods will
remain important in accreting pulsars studies even after the completion of the
Gaia mission, as reasonably accurate distance estimates can be
obtained even without precision astrometry, especially if several methods are
used and compared with each other. This is of particular importance for objects
located beyond $\sim5-10$\,kpc for which the accuracy of astrometric distances
is significantly degraded.

\begin{acknowledgements}
VD and AS thank the Deutsches Zentrum for Luft- und Raumfahrt (DLR)
and Deutsche Forschungsgemeinschaft (DFG) for financial support. \\

This work has made use of data from the European Space Agency (ESA) mission
{\it Gaia} (\url{https://www.cosmos.esa.int/gaia}), processed by the {\it Gaia}
Data Processing and Analysis Consortium (DPAC,
\url{https://www.cosmos.esa.int/web/gaia/dpac/consortium}). Funding for the DPAC
has been provided by national institutions, in particular the institutions
participating in the {\it Gaia} Multilateral Agreement.
\end{acknowledgements}

\bibliography{bibtex}   

\begin{thebibliography}{97}
\expandafter\ifx\csname natexlab\endcsname\relax\def\natexlab#1{#1}\fi

\bibitem[{{Aab} {et~al.}(1983){Aab}, {Bychkova}, \&
  {Kopylov}}]{1983SvAL....9..285A}
{Aab}, O.~E., {Bychkova}, L.~V., \& {Kopylov}, I.~M. 1983, Soviet Astronomy
  Letters, 9, 285

\bibitem[{{Ankay} {et~al.}(2001){Ankay}, {Kaper}, {de Bruijne}, {Dewi},
  {Hoogerwerf}, \& {Savonije}}]{2001AandA...370..170A}
{Ankay}, A., {Kaper}, L., {de Bruijne}, J.~H.~J., {et~al.} 2001, \aap, 370, 170

\bibitem[{{Bailer-Jones}(2015)}]{2015PASP..127..994B}
{Bailer-Jones}, C. A.~L. 2015, Publications of the Astronomical Society of the
  Pacific, 127, 994

\bibitem[{{Bailer-Jones} {et~al.}(2018){Bailer-Jones}, {Rybizki}, {Fouesneau},
  {Mantelet}, \& {Andrae}}]{2018arXiv180410121B}
{Bailer-Jones}, C.~A.~L., {Rybizki}, J., {Fouesneau}, M., {Mantelet}, G., \&
  {Andrae}, R. 2018, ArXiv e-prints, arXiv:1804.10121

\bibitem[{{Basko} {et~al.}(1975){Basko}, {Syunyaev}, \&
  {Sunyaev}}]{1975AandA....42..311B}
{Basko}, M.~M., {Syunyaev}, R.~A., \& {Sunyaev}, R.~A. 1975, \aap, 42, 311

\bibitem[{{Becker} {et~al.}(2012){Becker}, {Klochkov}, {Sch{\"o}nherr},
  {Nishimura}, {Ferrigno}, {Caballero}, {Kretschmar}, {Wolff}, {Wilms}, \&
  {Staubert}}]{2012AandA...544A.123B}
{Becker}, P.~A., {Klochkov}, D., {Sch{\"o}nherr}, G., {et~al.} 2012, \aap, 544

\bibitem[{{Bikmaev} {et~al.}(2017){Bikmaev}, {Shimansky}, {Irtuganov},
  {Glushkov}, {Sakhibullin}, {Khamitov}, {Burenin}, {Lutovinov}, {Zaznobin},
  {Pavlinsky}, {Sunyaev}, {Dodonov}, {Afanasiev}, {Kotov}, {Doroshenko}, \&
  {Tsygankov}}]{2017ATel10968....1B}
{Bikmaev}, I., {Shimansky}, V., {Irtuganov}, E., {et~al.} 2017, The
  Astronomer's Telegram, 10968

\bibitem[{{Blay} {et~al.}(2006){Blay}, {Negueruela}, {Reig}, {Coe}, {Corbet},
  {Fabregat}, \& {Tarasov}}]{2006AandA...446.1095B}
{Blay}, P., {Negueruela}, I., {Reig}, P., {et~al.} 2006, \aap, 446, 1095

\bibitem[{{Bolton} \& {Herbst}(1976)}]{1976AJ.....81..339B}
{Bolton}, C.~T. \& {Herbst}, W. 1976, \aj, 81, 339

\bibitem[{{Bonnet-Bidaud} \& {Mouchet}(1998)}]{1998AandA...332L...9B}
{Bonnet-Bidaud}, J.~M. \& {Mouchet}, M. 1998, \aap, 332, L9

\bibitem[{{Brucato} \& {Kristian}(1972)}]{1972ApJ...173L.105B}
{Brucato}, R.~J. \& {Kristian}, J. 1972, \apjl, 173, L105

\bibitem[{{Caballero} {et~al.}(2009){Caballero}, {Kraus}, {Postnov},
  {Santangelo}, {Kretschmar}, {Klochkov}, \& {Staubert}}]{2009arXiv0912.0252C}
{Caballero}, I., {Kraus}, U., {Postnov}, K., {et~al.} 2009, ArXiv e-prints

\bibitem[{{Chakrabarty} \& {Roche}(1997)}]{1997ApJ...489..254C}
{Chakrabarty}, D. \& {Roche}, P. 1997, \apj, 489, 254

\bibitem[{{Chevalier} \& {Ilovaisky}(1998)}]{1998AandA...330..201C}
{Chevalier}, C. \& {Ilovaisky}, S.~A. 1998, \aap, 330, 201

\bibitem[{{Clark}(2004)}]{2004ApJ...610..956C}
{Clark}, G.~W. 2004, \apj, 610, 956

\bibitem[{{Coe} \& {Payne}(1985)}]{1985ApandSS.109..175C}
{Coe}, M.~J. \& {Payne}, B.~J. 1985, \apss, 109, 175

\bibitem[{{Coleiro} \& {Chaty}(2013)}]{2013ApJ...764..185C}
{Coleiro}, A. \& {Chaty}, S. 2013, \apj, 764, 185

\bibitem[{{Cox} {et~al.}(2005){Cox}, {Kaper}, \&
  {Mokiem}}]{2005AandA...436..661C}
{Cox}, N.~L.~J., {Kaper}, L., \& {Mokiem}, M.~R. 2005, \aap, 436, 661

\bibitem[{{Day} \& {Tennant}(1991)}]{1991MNRAS.251...76D}
{Day}, C.~S.~R. \& {Tennant}, A.~F. 1991, \mnras, 251, 76

\bibitem[{{Doroshenko} {et~al.}(2017{\natexlab{a}}){Doroshenko}, {Tsygankov},
  \& {Santangelo}}]{2017arXiv171010912D}
{Doroshenko}, V., {Tsygankov}, S., \& {Santangelo}, A. 2017{\natexlab{a}},
  ArXiv e-prints

\bibitem[{{Doroshenko} {et~al.}(2017{\natexlab{b}}){Doroshenko}, {Tsygankov},
  {Mushtukov}, {Lutovinov}, {Santangelo}, {Suleimanov}, \&
  {Poutanen}}]{Doroshenko17}
{Doroshenko}, V., {Tsygankov}, S.~S., {Mushtukov}, A.~A., {et~al.}
  2017{\natexlab{b}}, \mnras, 466, 2143

\bibitem[{{Fabregat} {et~al.}(1992){Fabregat}, {Reglero}, {Coe}, {Clement},
  {Gorrod}, {Norton}, {Roche}, {Suso}, \& {Unger}}]{1992AandA...259..522F}
{Fabregat}, J., {Reglero}, V., {Coe}, M.~J., {et~al.} 1992, \aap, 259, 522

\bibitem[{{F{\"u}rst} {et~al.}(2014){F{\"u}rst}, {Pottschmidt}, {Wilms},
  {Tomsick}, {Bachetti}, {Boggs}, {Christensen}, {Craig}, {Grefenstette},
  {Hailey}, {Harrison}, {Madsen}, {Miller}, {Stern}, {Walton}, \&
  {Zhang}}]{2014ApJ...780..133F}
{F{\"u}rst}, F., {Pottschmidt}, K., {Wilms}, J., {et~al.} 2014, \apj, 780

\bibitem[{{Gaia Collaboration} {et~al.}(2018){Gaia Collaboration}, {Brown},
  {Vallenari}, {Prusti}, {de Bruijne}, {Babusiaux}, \&
  {Bailer-Jones}}]{2018arXiv180409365G}
{Gaia Collaboration}, {Brown}, A.~G.~A., {Vallenari}, A., {et~al.} 2018, ArXiv
  e-prints, arXiv:1804.09365

\bibitem[{{Gaia Collaboration} {et~al.}(2016){Gaia Collaboration}, {Prusti},
  {de Bruijne}, {Brown}, {Vallenari}, {Babusiaux}, {Bailer-Jones}, {Bastian},
  {Biermann}, {Evans}, {Eyer}, {Jansen}, {Jordi}, {Klioner}, {Lammers},
  {Lindegren}, {Luri}, {Mignard}, {Milligan}, {Panem}, {Poinsignon},
  {Pourbaix}, {Randich}, {Sarri}, {Sartoretti}, {Siddiqui}, {Soubiran},
  {Valette}, {van Leeuwen}, {Walton}, {Aerts}, {Arenou}, {Cropper}, {Drimmel},
  {H{\o}g}, {Katz}, {Lattanzi}, {O'Mullane}, {Grebel}, {Holland}, {Huc},
  {Passot}, {Bramante}, {Cacciari}, {Casta{\~n}eda}, {Chaoul}, {Cheek}, {De
  Angeli}, {Fabricius}, {Guerra}, {Hern{\'a}ndez}, {Jean-Antoine-Piccolo},
  {Masana}, {Messineo}, {Mowlavi}, {Nienartowicz}, {Ord{\'o}{\~n}ez- Blanco},
  {Panuzzo}, {Portell}, {Richards}, {Riello}, {Seabroke}, {Tanga},
  {Th{\'e}venin}, {Torra}, {Els}, {Gracia- Abril}, {Comoretto},
  {Garcia-Reinaldos}, {Lock}, {Mercier}, {Altmann}, {Andrae}, {Astraatmadja},
  {Bellas-Velidis}, {Benson}, {Berthier}, {Blomme}, {Busso}, {Carry},
  {Cellino}, {Clementini}, {Cowell}, {Creevey}, {Cuypers}, {Davidson}, {De
  Ridder}, {de Torres}, {Delchambre}, {Dell'Oro}, {Ducourant}, {Fr{\'e}mat},
  {Garc{\'\i}a-Torres}, {Gosset}, {Halbwachs}, {Hambly}, {Harrison}, {Hauser},
  {Hestroffer}, {Hodgkin}, {Huckle}, {Hutton}, {Jasniewicz}, {Jordan},
  {Kontizas}, {Korn}, {Lanzafame}, {Manteiga}, {Moitinho}, {Muinonen},
  {Osinde}, {Pancino}, {Pauwels}, {Petit}, {Recio-Blanco}, {Robin}, {Sarro},
  {Siopis}, {Smith}, {Smith}, {Sozzetti}, {Thuillot}, {van Reeven}, {Viala},
  {Abbas}, {Abreu Aramburu}, {Accart}, {Aguado}, {Allan}, {Allasia},
  {Altavilla}, {{\'A}lvarez}, {Alves}, {Anderson}, {Andrei}, {Anglada Varela},
  {Antiche}, {Antoja}, {Ant{\'o}n}, {Arcay}, {Atzei}, {Ayache}, {Bach},
  {Baker}, {Balaguer-N{\'u}{\~n}ez}, {Barache}, {Barata}, {Barbier}, {Barblan},
  {Baroni}, {Barrado y Navascu{\'e}s}, {Barros}, {Barstow}, {Becciani},
  {Bellazzini}, {Bellei}, {Bello Garc{\'\i}a}, {Belokurov}, {Bendjoya},
  {Berihuete}, {Bianchi}, {Bienaym{\'e}}, {Billebaud}, {Blagorodnova},
  {Blanco-Cuaresma}, {Boch}, {Bombrun}, {Borrachero}, {Bouquillon}, {Bourda},
  {Bouy}, {Bragaglia}, {Breddels}, {Brouillet}, {Br{\"u}semeister},
  {Bucciarelli}, {Budnik}, {Burgess}, {Burgon}, {Burlacu}, {Busonero}, {Buzzi},
  {Caffau}, {Cambras}, {Campbell}, {Cancelliere}, {Cantat-Gaudin}, {Carlucci},
  {Carrasco}, {Castellani}, {Charlot}, {Charnas}, {Charvet}, {Chassat},
  {Chiavassa}, {Clotet}, {Cocozza}, {Collins}, {Collins}, {Costigan}, {Crifo},
  {Cross}, {Crosta}, {Crowley}, {Dafonte}, {Damerdji}, {Dapergolas}, {David},
  {David}, {De Cat}, {de Felice}, {de Laverny}, {De Luise}, {De March}, {de
  Martino}, {de Souza}, {Debosscher}, {del Pozo}, {Delbo}, {Delgado},
  {Delgado}, {di Marco}, {Di Matteo}, {Diakite}, {Distefano}, {Dolding}, {Dos
  Anjos}, {Drazinos}, {Dur{\'a}n}, {Dzigan}, {Ecale}, {Edvardsson}, {Enke},
  {Erdmann}, {Escolar}, {Espina}, {Evans}, {Eynard Bontemps}, {Fabre},
  {Fabrizio}, {Faigler}, {Falc{\~a}o}, {Farr{\`a}s Casas}, {Faye}, {Federici},
  {Fedorets}, {Fern{\'a}ndez-Hern{\'a}ndez}, {Fernique}, {Fienga}, {Figueras},
  {Filippi}, {Findeisen}, {Fonti}, {Fouesneau}, {Fraile}, {Fraser}, {Fuchs},
  {Furnell}, {Gai}, {Galleti}, {Galluccio}, {Garabato}, {Garc{\'\i}a-Sedano},
  {Gar{\'e}}, {Garofalo}, {Garralda}, {Gavras}, {Gerssen}, {Geyer}, {Gilmore},
  {Girona}, {Giuffrida}, {Gomes}, {Gonz{\'a}lez-Marcos},
  {Gonz{\'a}lez-N{\'u}{\~n}ez}, {Gonz{\'a}lez-Vidal}, {Granvik}, {Guerrier},
  {Guillout}, {Guiraud}, {G{\'u}rpide}, {Guti{\'e}rrez-S{\'a}nchez}, {Guy},
  {Haigron}, {Hatzidimitriou}, {Haywood}, {Heiter}, {Helmi}, {Hobbs},
  {Hofmann}, {Holl}, {Holland}, {Hunt}, {Hypki}, {Icardi}, {Irwin}, {Jevardat
  de Fombelle}, {Jofr{\'e}}, {Jonker}, {Jorissen}, {Julbe}, {Karampelas},
  {Kochoska}, {Kohley}, {Kolenberg}, {Kontizas}, {Koposov}, {Kordopatis},
  {Koubsky}, {Kowalczyk}, {Krone-Martins}, {Kudryashova}, {Kull}, {Bachchan},
  {Lacoste-Seris}, {Lanza}, {Lavigne}, {Le Poncin-Lafitte}, {Lebreton},
  {Lebzelter}, {Leccia}, {Leclerc}, {Lecoeur-Taibi}, {Lemaitre}, {Lenhardt},
  {Leroux}, {Liao}, {Licata}, {Lindstr{\o}m}, {Lister}, {Livanou}, {Lobel},
  {L{\"o}ffler}, {L{\'o}pez}, {Lopez-Lozano}, {Lorenz}, {Loureiro},
  {MacDonald}, {Magalh{\~a}es Fernandes}, {Managau}, {Mann}, {Mantelet},
  {Marchal}, {Marchant}, {Marconi}, {Marie}, {Marinoni}, {Marrese},
  {Marschalk{\'o}}, {Marshall}, {Mart{\'\i}n-Fleitas}, {Martino}, {Mary},
  {Matijevi{\v{c}}}, {Mazeh}, {McMillan}, {Messina}, {Mestre}, {Michalik},
  {Millar}, {Miranda}, {Molina}, {Molinaro}, {Molinaro}, {Moln{\'a}r},
  {Moniez}, {Montegriffo}, {Monteiro}, {Mor}, {Mora}, {Morbidelli}, {Morel},
  {Morgenthaler}, {Morley}, {Morris}, {Mulone}, {Muraveva}, {Musella},
  {Narbonne}, {Nelemans}, {Nicastro}, {Noval}, {Ord{\'e}novic},
  {Ordieres-Mer{\'e}}, {Osborne}, {Pagani}, {Pagano}, {Pailler}, {Palacin},
  {Palaversa}, {Parsons}, {Paulsen}, {Pecoraro}, {Pedrosa}, {Pentik{\"a}inen},
  {Pereira}, {Pichon}, {Piersimoni}, {Pineau}, {Plachy}, {Plum}, {Poujoulet},
  {Pr{\v{s}}a}, {Pulone}, {Ragaini}, {Rago}, {Rambaux}, {Ramos-Lerate},
  {Ranalli}, {Rauw}, {Read}, {Regibo}, {Renk}, {Reyl{\'e}}, {Ribeiro},
  {Rimoldini}, {Ripepi}, {Riva}, {Rixon}, {Roelens}, {Romero-G{\'o}mez},
  {Rowell}, {Royer}, {Rudolph}, {Ruiz-Dern}, {Sadowski}, {Sagrist{\`a}
  Sell{\'e}s}, {Sahlmann}, {Salgado}, {Salguero}, {Sarasso}, {Savietto},
  {Schnorhk}, {Schultheis}, {Sciacca}, {Segol}, {Segovia}, {Segransan},
  {Serpell}, {Shih}, {Smareglia}, {Smart}, {Smith}, {Solano}, {Solitro},
  {Sordo}, {Soria Nieto}, {Souchay}, {Spagna}, {Spoto}, {Stampa}, {Steele},
  {Steidelm{\"u}ller}, {Stephenson}, {Stoev}, {Suess}, {S{\"u}veges}, {Surdej},
  {Szabados}, {Szegedi-Elek}, {Tapiador}, {Taris}, {Tauran}, {Taylor},
  {Teixeira}, {Terrett}, {Tingley}, {Trager}, {Turon}, {Ulla}, {Utrilla},
  {Valentini}, {van Elteren}, {Van Hemelryck}, {van Leeuwen}, {Varadi},
  {Vecchiato}, {Veljanoski}, {Via}, {Vicente}, {Vogt}, {Voss}, {Votruba},
  {Voutsinas}, {Walmsley}, {Weiler}, {Weingrill}, {Werner}, {Wevers},
  {Whitehead}, {Wyrzykowski}, {Yoldas}, {{\v{Z}}erjal}, {Zucker}, {Zurbach},
  {Zwitter}, {Alecu}, {Allen}, {Allende Prieto}, {Amorim},
  {Anglada-Escud{\'e}}, {Arsenijevic}, {Azaz}, {Balm}, {Beck}, {Bernstein},
  {Bigot}, {Bijaoui}, {Blasco}, {Bonfigli}, {Bono}, {Boudreault}, {Bressan},
  {Brown}, {Brunet}, {Bunclark}, {Buonanno}, {Butkevich}, {Carret}, {Carrion},
  {Chemin}, {Ch{\'e}reau}, {Corcione}, {Darmigny}, {de Boer}, {de Teodoro}, {de
  Zeeuw}, {Delle Luche}, {Domingues}, {Dubath}, {Fodor}, {Fr{\'e}zouls},
  {Fries}, {Fustes}, {Fyfe}, {Gallardo}, {Gallegos}, {Gardiol}, {Gebran},
  {Gomboc}, {G{\'o}mez}, {Grux}, {Gueguen}, {Heyrovsky}, {Hoar}, {Iannicola},
  {Isasi Parache}, {Janotto}, {Joliet}, {Jonckheere}, {Keil}, {Kim},
  {Klagyivik}, {Klar}, {Knude}, {Kochukhov}, {Kolka}, {Kos}, {Kutka}, {Lainey},
  {LeBouquin}, {Liu}, {Loreggia}, {Makarov}, {Marseille}, {Martayan},
  {Martinez-Rubi}, {Massart}, {Meynadier}, {Mignot}, {Munari}, {Nguyen},
  {Nordlander}, {Ocvirk}, {O'Flaherty}, {Olias Sanz}, {Ortiz}, {Osorio},
  {Oszkiewicz}, {Ouzounis}, {Palmer}, {Park}, {Pasquato}, {Peltzer}, {Peralta},
  {P{\'e}turaud}, {Pieniluoma}, {Pigozzi}, {Poels}, {Prat}, {Prod'homme},
  {Raison}, {Rebordao}, {Risquez}, {Rocca-Volmerange}, {Rosen}, {Ruiz-Fuertes},
  {Russo}, {Sembay}, {Serraller Vizcaino}, {Short}, {Siebert}, {Silva},
  {Sinachopoulos}, {Slezak}, {Soffel}, {Sosnowska}, {Strai{\v{z}}ys}, {ter
  Linden}, {Terrell}, {Theil}, {Tiede}, {Troisi}, {Tsalmantza}, {Tur},
  {Vaccari}, {Vachier}, {Valles}, {Van Hamme}, {Veltz}, {Virtanen}, {Wallut},
  {Wichmann}, {Wilkinson}, {Ziaeepour}, \& {Zschocke}}]{2016AandA...595A...1G}
{Gaia Collaboration}, {Prusti}, T., {de Bruijne}, J.~H.~J., {et~al.} 2016,
  \aap, 595, A1

\bibitem[{{Giangrande} {et~al.}(1980){Giangrande}, {Giovannelli}, {Bartolini},
  {Guarnieri}, \& {Piccioni}}]{1980AandAS...40..289G}
{Giangrande}, A., {Giovannelli}, F., {Bartolini}, C., {Guarnieri}, A., \&
  {Piccioni}, A. 1980, \aaps, 40, 289

\bibitem[{{Giangrande} {et~al.}(1977){Giangrande}, {Giovannelli}, {Bartolini},
  {Guranieri}, \& {Piccioni}}]{1977IAUC.3129....3G}
{Giangrande}, A., {Giovannelli}, F., {Bartolini}, C., {Guranieri}, A., \&
  {Piccioni}, A. 1977, \iaucirc, 3129

\bibitem[{{Gim{\'e}nez-Garc{\'{\i}}a}
  {et~al.}(2016){Gim{\'e}nez-Garc{\'{\i}}a}, {Shenar}, {Torrej{\'o}n},
  {Oskinova}, {Mart{\'{\i}}nez-N{\'u}{\~n}ez}, {Hamann}, {Rodes-Roca},
  {Gonz{\'a}lez-Gal{\'a}n}, {Alonso-Santiago}, {Gonz{\'a}lez-Fern{\'a}ndez},
  {Bernabeu}, \& {Sander}}]{2016AandA...591A..26G}
{Gim{\'e}nez-Garc{\'{\i}}a}, A., {Shenar}, T., {Torrej{\'o}n}, J.~M., {et~al.}
  2016, \aap, 591, A26

\bibitem[{{Hiltner} {et~al.}(1972){Hiltner}, {Werner}, \&
  {Osmer}}]{1972ApJ...175L..19H}
{Hiltner}, W.~A., {Werner}, J., \& {Osmer}, P. 1972, \apjl, 175, L19

\bibitem[{{Hinkle} {et~al.}(2006){Hinkle}, {Fekel}, {Joyce}, {Wood}, {Smith},
  \& {Lebzelter}}]{2006ApJ...641..479H}
{Hinkle}, K.~H., {Fekel}, F.~C., {Joyce}, R.~R., {et~al.} 2006, \apj, 641, 479

\bibitem[{{Honeycutt} \& {Schlegel}(1985)}]{1985PASP...97..300H}
{Honeycutt}, R.~K. \& {Schlegel}, E.~M. 1985, \pasp, 97, 300

\bibitem[{{Howarth} \& {Wilson}(1983)}]{1983MNRAS.202..347H}
{Howarth}, I.~D. \& {Wilson}, B. 1983, \mnras, 202, 347

\bibitem[{{Hutchings} {et~al.}(1978){Hutchings}, {Bernard}, {Crampton}, \&
  {Cowley}}]{1978ApJ...223..530H}
{Hutchings}, J.~B., {Bernard}, J.~E., {Crampton}, D., \& {Cowley}, A.~P. 1978,
  \apj, 223, 530

\bibitem[{{Hutchings} {et~al.}(1979){Hutchings}, {Cowley}, {Crampton}, {van
  Paradijs}, \& {White}}]{1979ApJ...229.1079H}
{Hutchings}, J.~B., {Cowley}, A.~P., {Crampton}, D., {van Paradijs}, J., \&
  {White}, N.~E. 1979, \apj, 229, 1079

\bibitem[{{I{\c{c}}dem} {et~al.}(2011){I{\c{c}}dem}, {Inam}, \&
  {Baykal}}]{2011MNRAS.415.1523I}
{I{\c{c}}dem}, B., {Inam}, S., \& {Baykal}, A. 2011, \mnras, 415, 1523

\bibitem[{{Ilovaisky} {et~al.}(1979){Ilovaisky}, {Chevalier}, \&
  {Motch}}]{1979AandA....71L..17I}
{Ilovaisky}, S.~A., {Chevalier}, C., \& {Motch}, C. 1979, \aap, 71, L17

\bibitem[{{Israel} {et~al.}(2001){Israel}, {Negueruela}, {Campana}, {Covino},
  {Di Paola}, {Maxwell}, {Norton}, {Speziali}, {Verrecchia}, \&
  {Stella}}]{2001AandA...371.1018I}
{Israel}, G.~L., {Negueruela}, I., {Campana}, S., {et~al.} 2001, \aap, 371,
  1018

\bibitem[{{Janot-Pacheco} {et~al.}(1981){Janot-Pacheco}, {Ilovaisky}, \&
  {Chevalier}}]{1981AandA....99..274J}
{Janot-Pacheco}, E., {Ilovaisky}, S.~A., \& {Chevalier}, C. 1981, \aap, 99, 274

\bibitem[{{Janot-Pacheco} {et~al.}(1987){Janot-Pacheco}, {Motch}, \&
  {Mouchet}}]{1987AandA...177...91J}
{Janot-Pacheco}, E., {Motch}, C., \& {Mouchet}, M. 1987, \aap, 177, 91

\bibitem[{{Kaper} {et~al.}(1995){Kaper}, {Lamers}, {Ruymaekers}, {van den
  Heuvel}, \& {Zuiderwijk}}]{1995AandA...300..446K}
{Kaper}, L., {Lamers}, H.~J.~G.~L.~M., {Ruymaekers}, E., {van den Heuvel},
  E.~P.~J., \& {Zuiderwijk}, E.~J. 1995, \aap, 300

\bibitem[{{Kaper} {et~al.}(2006){Kaper}, {van der Meer}, \&
  {Najarro}}]{2006AandA...457..595K}
{Kaper}, L., {van der Meer}, A., \& {Najarro}, F. 2006, \aap, 457, 595

\bibitem[{{Koenigsberger} {et~al.}(2003){Koenigsberger}, {Canalizo}, {Arrieta},
  {Richer}, \& {Georgiev}}]{2003RMxAA..39...17K}
{Koenigsberger}, G., {Canalizo}, G., {Arrieta}, A., {Richer}, M.~G., \&
  {Georgiev}, L. 2003, \rmxaa, 39, 17

\bibitem[{{Krzeminski}(1974)}]{1974ApJ...192L.135K}
{Krzeminski}, W. 1974, \apjl, 192, L135

\bibitem[{{Leahy} \& {Abdallah}(2014)}]{2014ApJ...793...79L}
{Leahy}, D.~A. \& {Abdallah}, M.~H. 2014, \apj, 793, 79

\bibitem[{{Lipunov}(1981)}]{Lipunov}
{Lipunov}, V.~M. 1981, \sovast, 25, 375

\bibitem[{{Luri} {et~al.}(2018){Luri}, {Brown}, {Sarro}, {Arenou},
  {Bailer-Jones}, {Castro-Ginard}, {de Bruijne}, {Prusti}, {Babusiaux}, \&
  {Delgado}}]{2018arXiv180409376L}
{Luri}, X., {Brown}, A.~G.~A., {Sarro}, L.~M., {et~al.} 2018, ArXiv e-prints,
  arXiv:1804.09376

\bibitem[{{Lyubimkov} {et~al.}(1997){Lyubimkov}, {Rostopchin}, {Roche}, \&
  {Tarasov}}]{1997MNRAS.286..549L}
{Lyubimkov}, L.~S., {Rostopchin}, S.~I., {Roche}, P., \& {Tarasov}, A.~E. 1997,
  \mnras, 286, 549

\bibitem[{{Mart{\'\i}nez-N{\'u}{\~n}ez}
  {et~al.}(2015){Mart{\'\i}nez-N{\'u}{\~n}ez}, {Sander}, {G{\'\i}menez-
  Garc{\'\i}a}, {G{\'o}nzalez-Gal{\'a}n}, {Torrej{\'o}n},
  {G{\'o}nzalez-Fern{\'a}ndez}, \& {Hamann}}]{2015AandA...578A.107M}
{Mart{\'\i}nez-N{\'u}{\~n}ez}, S., {Sander}, A., {G{\'\i}menez- Garc{\'\i}a},
  A., {et~al.} 2015, \aap, 578, A107

\bibitem[{{Mason} \& {Cordova}(1982)}]{1982APJ...262..253M}
{Mason}, K.~O. \& {Cordova}, F.~A. 1982, \apj, 262, 253

\bibitem[{{Mason} {et~al.}(1978){Mason}, {Murdin}, {Parkes}, \&
  {Visvanathan}}]{1978MNRAS.184P..45M}
{Mason}, K.~O., {Murdin}, P.~G., {Parkes}, G.~E., \& {Visvanathan}, N. 1978,
  \mnras, 184, 45P

\bibitem[{{McBride} {et~al.}(2006){McBride}, {Wilms}, {Coe}, {Kreykenbohm},
  {Rothschild}, {Coburn}, {Galache}, {Kretschmar}, {Edge}, \&
  {Staubert}}]{2006AandA...451..267M}
{McBride}, V.~A., {Wilms}, J., {Coe}, M.~J., {et~al.} 2006, \aap, 451, 267

\bibitem[{{Megier} {et~al.}(2009){Megier}, {Strobel}, {Galazutdinov}, \&
  {Kre{\l}owski}}]{2009AandA...507..833M}
{Megier}, A., {Strobel}, A., {Galazutdinov}, G.~A., \& {Kre{\l}owski}, J. 2009,
  \aap, 507, 833

\bibitem[{{Menzies}(1981)}]{1981MNRAS.195P..67M}
{Menzies}, J. 1981, \mnras, 195, 67P

\bibitem[{{Mihara} {et~al.}(1998){Mihara}, {Makishima}, \&
  {Nagase}}]{1998AdSpR..22..987M}
{Mihara}, T., {Makishima}, K., \& {Nagase}, F. 1998, Advances in Space
  Research, 22, 987

\bibitem[{{Morel} \& {Grosdidier}(2005)}]{2005MNRAS.356..665M}
{Morel}, T. \& {Grosdidier}, Y. 2005, \mnras, 356, 665

\bibitem[{{Motch} {et~al.}(1997){Motch}, {Haberl}, {Dennerl}, {Pakull}, \&
  {Janot-Pacheco}}]{1997AandA...323..853M}
{Motch}, C., {Haberl}, F., {Dennerl}, K., {Pakull}, M., \& {Janot-Pacheco}, E.
  1997, \aap, 323, 853

\bibitem[{{Mushtukov} {et~al.}(2015{\natexlab{a}}){Mushtukov}, {Suleimanov},
  {Tsygankov}, \& {Poutanen}}]{Mushtukov15}
{Mushtukov}, A.~A., {Suleimanov}, V.~F., {Tsygankov}, S.~S., \& {Poutanen}, J.
  2015{\natexlab{a}}, \mnras, 454, 2539

\bibitem[{{Mushtukov} {et~al.}(2015{\natexlab{b}}){Mushtukov}, {Suleimanov},
  {Tsygankov}, \& {Poutanen}}]{2015MNRAS.447.1847M}
{Mushtukov}, A.~A., {Suleimanov}, V.~F., {Tsygankov}, S.~S., \& {Poutanen}, J.
  2015{\natexlab{b}}, \mnras, 447, 1847

\bibitem[{{Negueruela} {et~al.}(2003){Negueruela}, {Israel}, {Marco}, {Norton},
  \& {Speziali}}]{2003AandA...397..739N}
{Negueruela}, I., {Israel}, G.~L., {Marco}, A., {Norton}, A.~J., \& {Speziali},
  R. 2003, \aap, 397, 739

\bibitem[{{Negueruela} \& {Okazaki}(2001)}]{2001AandA...369..108N}
{Negueruela}, I. \& {Okazaki}, A.~T. 2001, \aap, 369, 108

\bibitem[{{Negueruela} \& {Reig}(2001)}]{2001AandA...371.1056N}
{Negueruela}, I. \& {Reig}, P. 2001, \aap, 371, 1056

\bibitem[{{Negueruela} {et~al.}(1999){Negueruela}, {Roche}, {Fabregat}, \&
  {Coe}}]{1999MNRAS.307..695N}
{Negueruela}, I., {Roche}, P., {Fabregat}, J., \& {Coe}, M.~J. 1999, \mnras,
  307, 695

\bibitem[{{Nespoli} {et~al.}(2008){Nespoli}, {Fabregat}, \&
  {Mennickent}}]{2008AandA...486..911N}
{Nespoli}, E., {Fabregat}, J., \& {Mennickent}, R.~E. 2008, \aap, 486, 911

\bibitem[{{Parkes} {et~al.}(1980{\natexlab{a}}){Parkes}, {Mason}, {Murdin}, \&
  {Culhane}}]{1980MNRAS.191..547P}
{Parkes}, G.~E., {Mason}, K.~O., {Murdin}, P.~G., \& {Culhane}, J.~L.
  1980{\natexlab{a}}, \mnras, 191, 547

\bibitem[{{Parkes} {et~al.}(1978){Parkes}, {Murdin}, \&
  {Mason}}]{1978MNRAS.184P..73P}
{Parkes}, G.~E., {Murdin}, P.~G., \& {Mason}, K.~O. 1978, \mnras, 184, 73P

\bibitem[{{Parkes} {et~al.}(1980{\natexlab{b}}){Parkes}, {Murdin}, \&
  {Mason}}]{1980MNRAS.190..537P}
{Parkes}, G.~E., {Murdin}, P.~G., \& {Mason}, K.~O. 1980{\natexlab{b}}, \mnras,
  190, 537

\bibitem[{{Parmar} {et~al.}(1989){Parmar}, {White}, {Stella}, {Izzo}, \&
  {Ferri}}]{1989ApJ...338..359P}
{Parmar}, A.~N., {White}, N.~E., {Stella}, L., {Izzo}, C., \& {Ferri}, P. 1989,
  \apj, 338, 359

\bibitem[{{Pellizza} {et~al.}(2006){Pellizza}, {Chaty}, \&
  {Negueruela}}]{2006AandA...455..653P}
{Pellizza}, L.~J., {Chaty}, S., \& {Negueruela}, I. 2006, \aap, 455, 653

\bibitem[{{Poutanen} {et~al.}(2013){Poutanen}, {Mushtukov}, {Suleimanov},
  {Tsygankov}, {Nagirner}, {Doroshenko}, \& {Lutovinov}}]{Poutanen13}
{Poutanen}, J., {Mushtukov}, A.~A., {Suleimanov}, V.~F., {et~al.} 2013, \apj,
  777, 115

\bibitem[{{Rahoui} {et~al.}(2008){Rahoui}, {Chaty}, {Lagage}, \&
  {Pantin}}]{2008AandA...484..801R}
{Rahoui}, F., {Chaty}, S., {Lagage}, P.-O., \& {Pantin}, E. 2008, \aap, 484,
  801

\bibitem[{{Rappaport} {et~al.}(1978){Rappaport}, {Clark}, {Cominsky}, {Joss},
  \& {Li}}]{1978ApJ...224L...1R}
{Rappaport}, S., {Clark}, G.~W., {Cominsky}, L., {Joss}, P.~C., \& {Li}, F.
  1978, \apjl, 224, L1

\bibitem[{{Reig} {et~al.}(1996){Reig}, {Chakrabarty}, {Coe}, {Fabregat},
  {Negueruela}, {Prince}, {Roche}, \& {Steele}}]{1996AandA...311..879R}
{Reig}, P., {Chakrabarty}, D., {Coe}, M.~J., {et~al.} 1996, \aap, 311, 879

\bibitem[{{Reig} \& {Fabregat}(2015)}]{2015AandA...574A..33R}
{Reig}, P. \& {Fabregat}, J. 2015, \aap, 574, A33

\bibitem[{{Reig} {et~al.}(2007){Reig}, {Larionov}, {Negueruela}, {Arkharov}, \&
  {Kudryavtseva}}]{2007AandA...462.1081R}
{Reig}, P., {Larionov}, V., {Negueruela}, I., {Arkharov}, A.~A., \&
  {Kudryavtseva}, N.~A. 2007, \aap, 462, 1081

\bibitem[{{Reig} {et~al.}(2005){Reig}, {Negueruela}, {Fabregat}, {Chato}, \&
  {Coe}}]{2005AandA...440.1079R}
{Reig}, P., {Negueruela}, I., {Fabregat}, J., {Chato}, R., \& {Coe}, M.~J.
  2005, \aap, 440, 1079

\bibitem[{{Reig} {et~al.}(2011){Reig}, {Nespoli}, {Fabregat}, \&
  {Mennickent}}]{2011AandA...533A..23R}
{Reig}, P., {Nespoli}, E., {Fabregat}, J., \& {Mennickent}, R.~E. 2011, \aap,
  533, A23

\bibitem[{{Reynolds} {et~al.}(1992){Reynolds}, {Bell}, \&
  {Hilditch}}]{1992MNRAS.256..631R}
{Reynolds}, A.~P., {Bell}, S.~A., \& {Hilditch}, R.~W. 1992, \mnras, 256, 631

\bibitem[{{Reynolds} {et~al.}(1997){Reynolds}, {Quaintrell}, {Still}, {Roche},
  {Chakrabarty}, \& {Levine}}]{1997MNRAS.288...43R}
{Reynolds}, A.~P., {Quaintrell}, H., {Still}, M.~D., {et~al.} 1997, \mnras,
  288, 43

\bibitem[{{Riquelme} {et~al.}(2012){Riquelme}, {Torrej{\'o}n}, \&
  {Negueruela}}]{2012AandA...539A.114R}
{Riquelme}, M.~S., {Torrej{\'o}n}, J.~M., \& {Negueruela}, I. 2012, \aap, 539,
  A114

\bibitem[{{Rodriguez} {et~al.}(2009){Rodriguez}, {Tomsick}, {Bodaghee}, {Zurita
  Heras}, {Chaty}, {Paizis}, \& {Corbel}}]{2009AandA...508..889R}
{Rodriguez}, J., {Tomsick}, J.~A., {Bodaghee}, A., {et~al.} 2009, \aap, 508,
  889

\bibitem[{{Sadakane} {et~al.}(1985){Sadakane}, {Hirata}, {Jugaku}, {Kondo},
  {Matsuoka}, {Tanaka}, \& {Hammerschlag-Hensberge}}]{1985ApJ...288..284S}
{Sadakane}, K., {Hirata}, R., {Jugaku}, J., {et~al.} 1985, \apj, 288, 284

\bibitem[{{Schwartz} {et~al.}(1980){Schwartz}, {Griffiths}, {Thorstensen},
  {Charles}, \& {Bowyer}}]{1980AJ.....85..549S}
{Schwartz}, D.~A., {Griffiths}, R.~E., {Thorstensen}, J.~R., {Charles}, P.~A.,
  \& {Bowyer}, S. 1980, \aj, 85, 549

\bibitem[{{Scott} {et~al.}(1997){Scott}, {Finger}, {Wilson}, {Koh}, {Prince},
  {Vaughan}, \& {Chakrabarty}}]{Scott97}
{Scott}, D.~M., {Finger}, M.~H., {Wilson}, R.~B., {et~al.} 1997, \apj, 488, 831

\bibitem[{{Takagi} {et~al.}(2016){Takagi}, {Mihara}, {Sugizaki}, {Makishima},
  \& {Morii}}]{2016PASJ...68S..13T}
{Takagi}, T., {Mihara}, T., {Sugizaki}, M., {Makishima}, K., \& {Morii}, M.
  2016, Publications of the Astronomical Society of Japan, 68, S13

\bibitem[{{Telting} {et~al.}(1998){Telting}, {Waters}, {Roche}, {Boogert},
  {Clark}, {de Martino}, \& {Persi}}]{1998MNRAS.296..785T}
{Telting}, J.~H., {Waters}, L.~B.~F.~M., {Roche}, P., {et~al.} 1998, \mnras,
  296, 785

\bibitem[{{Thompson} \& {Rothschild}(2009)}]{2009ApJ...691.1744T}
{Thompson}, T.~W.~J. \& {Rothschild}, R.~E. 2009, \apj, 691, 1744

\bibitem[{{Tr{\"u}mper} \& {Sch{\"o}nfelder}(1973)}]{1973AandA....25..445T}
{Tr{\"u}mper}, J. \& {Sch{\"o}nfelder}, V. 1973, \aap, 25, 445

\bibitem[{{Tsygankov} \& {Lutovinov}(2005)}]{2005AstL...31...88T}
{Tsygankov}, S.~S. \& {Lutovinov}, A.~A. 2005, Astronomy Letters, 31, 88

\bibitem[{{van Genderen} \& {Sterken}(1996)}]{1996AandA...308..763V}
{van Genderen}, A.~M. \& {Sterken}, C. 1996, \aap, 308, 763

\bibitem[{{van Oijen}(1989)}]{1989AandA...217..115V}
{van Oijen}, J.~G.~J. 1989, \aap, 217, 115

\bibitem[{{Verrecchia} {et~al.}(2002){Verrecchia}, {Israel}, {Negueruela},
  {Covino}, {Polcaro}, {Clark}, {Steele}, {Gualandi}, {Speziali}, \&
  {Stella}}]{2002AandA...393..983V}
{Verrecchia}, F., {Israel}, G.~L., {Negueruela}, I., {et~al.} 2002, \aap, 393,
  983

\bibitem[{{Vidal}(1973)}]{1973ApJ...186L..81V}
{Vidal}, N.~V. 1973, \apjl, 186, L81

\bibitem[{{Vybornov} {et~al.}(2018){Vybornov}, {Doroshenko}, {Staubert}, \&
  {Santangelo}}]{Vybornov18}
{Vybornov}, V., {Doroshenko}, V., {Staubert}, R., \& {Santangelo}, A. 2018,
  \aap, 610, A88

\bibitem[{{Wade} \& {Oke}(1977)}]{1977ApJ...215..568W}
{Wade}, R.~A. \& {Oke}, J.~B. 1977, \apj, 215, 568

\bibitem[{{Wilson} {et~al.}(2002){Wilson}, {Finger}, {Coe}, {Laycock}, \&
  {Fabregat}}]{2002ApJ...570..287W}
{Wilson}, C.~A., {Finger}, M.~H., {Coe}, M.~J., {Laycock}, S., \& {Fabregat},
  J. 2002, \apj, 570, 287

\bibitem[{{Wilson} {et~al.}(2003){Wilson}, {Finger}, {Coe}, \&
  {Negueruela}}]{2003ApJ...584..996W}
{Wilson}, C.~A., {Finger}, M.~H., {Coe}, M.~J., \& {Negueruela}, I. 2003, \apj,
  584, 996

\bibitem[{{Zuiderwijk} {et~al.}(1974){Zuiderwijk}, {van den Heuvel}, \&
  {Hensberge}}]{1974AandA....35..353Z}
{Zuiderwijk}, E.~J., {van den Heuvel}, E.~P.~J., \& {Hensberge}, G. 1974, \aap,
  35, 353

\end{thebibliography}

\onecolumn
\begin{longtable}{llllll}
\multicolumn{6}{L{19cm}}{ \small{\textbf{Table \ref{tab:src}.} Sample of X-ray pulsars considered in this work. The columns $d_{\mathrm px}$, $d_{\mathrm BJ}$, $d_{other}$ show the
distances obtained by parallax inversion, Bayesian analysis of Gaia DR2 data \citep{2018arXiv180410121B}, and 
    range of distances published in the literature respectively. The corresponding references are given in the last column. For each
    distance estimate the respective publication and method are given in brackets. Sources without a distance obtained by parallax inversion are those with measured negative parallax or parallax smaller than its error.}} \\ \\
 Source & Gaia DR2 ID${}^{82}$ & $d_{\mathrm px}{}^{82}$ [kpc]& $d_{\mathrm BJ}{}^{81}$ [kpc] & $d_{\mathrm other}$ [kpc] & $d_{\mathrm other}$ detail [kpc], Refs. \\ \hline 
\endfirsthead 

\multicolumn{6}{L{19cm}}{\small{\textbf{Table \ref{tab:src}.} Continued.}}\\
 Source & Gaia DR2 ID${}^{82}$ & $d_{\mathrm px}{}^{82}$ [kpc]& $d_{\mathrm BJ}{}^{81}$ [kpc] & $d_{\mathrm other}$ [kpc] & $d_{\mathrm other}$ detail [kpc], Refs. \\ \hline  \\
\hline
\endhead 

\endfoot
\hline
\multicolumn{6}{L{19cm}}{\small{
\textbf{References} $^{1}$-\cite{1972ApJ...173L.105B},
$^{2}$-\cite{1972ApJ...175L..19H},
$^{3}$-\cite{1973ApJ...186L..81V},
$^{4}$-\cite{1974AandA....35..353Z},
$^{5}$-\cite{1974ApJ...192L.135K},
$^{6}$-\cite{1976AJ.....81..339B},
$^{7}$-\cite{1977ApJ...215..568W},
$^{8}$-\cite{1977IAUC.3129....3G},
$^{9}$-\cite{1978ApJ...223..530H},
$^{10}$-\cite{1978ApJ...224L...1R},
$^{11}$-\cite{1978MNRAS.184P..45M},
$^{12}$-\cite{1978MNRAS.184P..73P},
$^{13}$-\cite{1979AandA....71L..17I},
$^{14}$-\cite{1979ApJ...229.1079H},
$^{15}$-\cite{1980AandAS...40..289G},
$^{16}$-\cite{1980AJ.....85..549S},
$^{17}$-\cite{1980MNRAS.190..537P},
$^{18}$-\cite{1980MNRAS.191..547P},
$^{19}$-\cite{1981AandA....99..274J},
$^{20}$-\cite{1981MNRAS.195P..67M},
$^{21}$-\cite{1982APJ...262..253M},
$^{22}$-\cite{1983MNRAS.202..347H},
$^{23}$-\cite{1983SvAL....9..285A},
$^{24}$-\cite{1985ApandSS.109..175C},
$^{25}$-\cite{1985ApJ...288..284S},
$^{26}$-\cite{1985PASP...97..300H},
$^{27}$-\cite{1987AandA...177...91J},
$^{28}$-\cite{1989AandA...217..115V},
$^{29}$-\cite{1989ApJ...338..359P},
$^{30}$-\cite{1991MNRAS.251...76D},
$^{31}$-\cite{1992AandA...259..522F},
$^{32}$-\cite{1992MNRAS.256..631R},
$^{33}$-\cite{1995AandA...300..446K},
$^{34}$-\cite{1996AandA...308..763V},
$^{35}$-\cite{1996AandA...311..879R},
$^{36}$-\cite{1997AandA...323..853M},
$^{37}$-\cite{1997ApJ...489..254C},
$^{38}$-\cite{1997MNRAS.286..549L},
$^{39}$-\cite{1997MNRAS.288...43R},
$^{40}$-\cite{1998AandA...330..201C},
$^{41}$-\cite{1998AandA...332L...9B},
$^{42}$-\cite{1998AdSpR..22..987M},
$^{43}$-\cite{1998MNRAS.296..785T},
$^{44}$-\cite{1999MNRAS.307..695N},
$^{45}$-\cite{2001AandA...369..108N},
$^{46}$-\cite{2001AandA...370..170A},
$^{47}$-\cite{2001AandA...371.1018I},
$^{48}$-\cite{2001AandA...371.1056N},
$^{49}$-\cite{2002AandA...393..983V},
$^{50}$-\cite{2002ApJ...570..287W},
$^{51}$-\cite{2003AandA...397..739N},
$^{52}$-\cite{2003ApJ...584..996W},
$^{53}$-\cite{2003RMxAA..39...17K},
$^{54}$-\cite{2004ApJ...610..956C},
$^{55}$-\cite{2005AandA...436..661C},
$^{56}$-\cite{2005AandA...440.1079R},
$^{57}$-\cite{2005AstL...31...88T},
$^{58}$-\cite{2005MNRAS.356..665M},
$^{59}$-\cite{2006AandA...446.1095B},
$^{60}$-\cite{2006AandA...451..267M},
$^{61}$-\cite{2006AandA...455..653P},
$^{62}$-\cite{2006AandA...457..595K},
$^{63}$-\cite{2006ApJ...641..479H},
$^{64}$-\cite{2007AandA...462.1081R},
$^{65}$-\cite{2008AandA...484..801R},
$^{66}$-\cite{2008AandA...486..911N},
$^{67}$-\cite{2009AandA...507..833M},
$^{68}$-\cite{2009AandA...508..889R},
$^{69}$-\cite{2009ApJ...691.1744T},
$^{70}$-\cite{2011AandA...533A..23R},
$^{71}$-\cite{2011MNRAS.415.1523I},
$^{72}$-\cite{2012AandA...539A.114R},
$^{73}$-\cite{2013ApJ...764..185C},
$^{74}$-\cite{2014ApJ...793...79L},
$^{75}$-\cite{2015AandA...574A..33R},
$^{76}$-\cite{2015AandA...578A.107M},
$^{77}$-\cite{2016AandA...591A..26G},
$^{78}$-\cite{2016PASJ...68S..13T},
$^{79}$-\cite{2017arXiv171010912D},
$^{80}$-\cite{2017ATel10968....1B},
$^{81}$-\cite{2018arXiv180410121B}
$^{82}$-\cite{2018arXiv180409365G}
}} \\

\endlastfoot

1A 0535+262 		& 3441207615-  & 2.26 $\pm$ 0.25 &$ 2.13 ^{+ 0.21 }_{- 0.26 }$& 0.2-4.6 &			$0.33^{+0,45}_{-0,12}$,PX$^{40}$;$3.8\pm0.33$, SED$^{73}$;$2.3^{+2,3}_{-0,8}$,SPH$^{8}$;		\\ & 229815040 &&&&$1.3$, SPH$^{9}$;$1.8\pm0.6$, SPH$^{15}$;$2.6\pm0.4$, SPH$^{27}$;			\\ &&&&&$3.9\pm0.1$, SPH$^{60}$;$2.1\pm0.5$, SPH$^{75}$\\

1A 1118-615 		& 5336957010-  & 3.19 $\pm$ 0.28 &$ 2.93 ^{+ 0.22 }_{- 0.26 }$& 1.8-7.0 &			$4$, ATM$^{24}$;$3.2\pm1.4$, SED$^{73}$;				\\ & 898124160 &&&&$5\pm2$, SPH$^{19}$;$5.2\pm0.9$, SPH$^{72}$\\

2A 1822-371 		& 6728016172-  & 8.30 $\pm$ 3.98 &$ 6.05 ^{+ 1.57 }_{- 2.68 }$& 1.0-5.0 &			$2.5^{+2,5}_{-1,5}$,LC$^{21}$						\\ & 687965568 &&&& \\

2S 0114+650 		& 5249243101-  & 9.60 $\pm$ 3.03 &$ 6.56 ^{+ 1.12 }_{- 1.57 }$& 1.5-10.6 &			$1.8\pm0.3$, SPH$^{23}$;$7\pm3.6$, SPH$^{35}$;$3.8$, SPH$^{40}$;				\\ &53249920&&&&$6.5\pm3$, SPH$^{73}$;$5.9\pm1.4$, SPH$^{75}$;$3$, SV$^{53}$\\

3A 1728-247 		& 4110236324-  &    &$ 7.61 ^{+ 2.76 }_{- 4.28 }$& 3.0-15.0 &					$4.3$, SPH$^{63}$;$9\pm6$, SU$^{37}$					\\ & 513030656 &&&& \\

4U 0115+634 		& 5246774697-  & 11.01 $\pm$ 3.26 &$ 7.20 ^{+ 1.10 }_{- 1.48 }$& 2.5-9.0 &			$8.1\pm0.1$, ATM$^{64}$;$5.3\pm0.44$, SED$^{73}$;$2.5$, SU$^{10}$			\\ &90488960&&&&$7\pm0.3$, SPH$^{72}$;$6\pm1.5$, SPH$^{75}$;$8\pm1$, SPH$^{45}$;			\\

4U 1538-52 		& 5886085557-  & 13.58 $\pm$ 9.18 &$ 6.65 ^{+ 1.44 }_{- 2.11 }$& 3.9-8.0 &			$4.5$, DH$^{54}$;$6.2\pm1.8$, SED$^{73}$;$5.5^{+2,3}_{-1,6}$,SPH$^{12}$;				\\ &746480000&&&&$6\pm0.5$, SPH$^{13}$;$6.4\pm1$, SPH$^{32}$		\\

4U 1626-67 		& 5809528276-  &   &$ 3.46 ^{+ 1.28 }_{- 2.28 }$& 5.0-13.0 &					$9\pm4$, SU$^{78}$							\\ & 749789312 &&&& \\

4U 1700-37 		& 5976382915-  & 1.82 $\pm$ 0.21 &$ 1.75 ^{+ 0.19 }_{- 0.23 }$& 1.4-2.0 &			$1.7\pm0.3$, NH$^{6}$;$1.9$, RA$^{46}$;					\\ &813535232&&&&$1.8\pm0.15$, SED$^{73}$\\

4U 1907+09  		& 4309225217-  & 7.44 $\pm$ 5.67 &$ 4.39 ^{+ 1.22 }_{- 2.01 }$& 1.0-13.0 &			$7.5\pm5.5$, SPH$^{16}$;$5$, SPH$^{55}$;				\\ &336729088&&&&$2.8^{+5}_{-1,8}$,SPH$^{66}$;$4.4\pm1.2$, SPH$^{75}$\\

4U 1909+07 		& 4306419980-  &    &$ 2.59 ^{+ 1.32 }_{- 2.25 }$& 4.0-10.0 &					$4.85\pm0.5$, ATM$^{76}$;$7\pm3$, SPH$^{58}$				\\ &916246656&&&&\\

4U 2206+54 		& 2005653524-  & 3.74 $\pm$ 0.43 &$ 3.34 ^{+ 0.32 }_{- 0.39 }$& 2.3-3.8 &			$3.4\pm0.35$, SED$^{73}$;$3$, SPH$^{48}$;				\\ &280214400&&&&$2.6$, SPH$^{59}$;$3\pm0.7$, SPH$^{75}$\\

Cen X-3 		& 5337498593-  & 8.70 $\pm$ 2.27 &$ 6.37 ^{+ 1.01 }_{- 1.39 }$& 4.2-10.0 &			$5.4\pm0.3$, DH$^{30}$;$5.7\pm1.5$, DH$^{69}$;				\\ &446516480&&&&$8_{-1.8}$,SPH$^{5}$;$10$, SPH$^{14}$\\

Cep X-4 		& 2178178409-  & 19.73 $\pm$ 7.93 &$ 10.17 ^{+ 1.59 }_{- 2.14 }$& 2.8-9.1 &			$3.2\pm0.4$, ACH$^{42}$;$3.7\pm0.52$, SED$^{73}$;			\\ &188167296&&&&$7.9\pm1.2$, SPH$^{72}$;$5.9\pm0.9$, SPH$^{72}$		\\ &&&&&$3.8\pm0.6$, SPH$^{41}$;\\

EXO 2030+375 		& 2063791369-  & 6.79 $\pm$ 4.92 &$ 3.64 ^{+ 0.88 }_{- 1.34 }$& 2.7-9.0 &			$7.1\pm0.3$, NH$^{50}$;$3.1\pm0.38$, SED$^{73}$;			\\ &815322752&&&&$6.5\pm2.5$, SPH$^{75}$;$5.3\pm0.3$, SU$^{29}$\\

GX 301-2 		& 6054569565-  & 3.96 $\pm$ 0.55 &$ 3.53 ^{+ 0.40 }_{- 0.52 }$& 1.4-10.0 &			$3$, ATM$^{62}$;$3.1\pm0.64$, SED$^{73}$;$2$, SPH$^{3}$;				\\ &614460800&&&&$1.8\pm0.4$, SPH$^{18}$;$5.3$, SPH$^{33}$;$7.7\pm2.3$, SPH$^{34}$				\\

GX 304-1 		& 5863533199-  & 2.13 $\pm$ 0.15 &$ 2.01 ^{+ 0.13 }_{- 0.15 }$& 1.0-3.0 &			$3$, NH$^{20}$;$1.3\pm0.1$, SED$^{73}$;					\\ &843070208&&&&$2\pm1$, SPH$^{11}$;$2.4\pm0.5$, SPH$^{17}$\\

Her X-1 		& 1338822021-  & 6.72 $\pm$ 1.20 &$ 5.04 ^{+ 0.57 }_{- 0.71 }$& 5.7-7.0 &			$6.1^{+0.88}_{-0.37}$,ATM$^{74}$;$5.95\pm0.05$, ATM$^{22}$;		\\ &487330304&&&&$6.6\pm0.4$, EB$^{39}$\\

IGR J17544-2619 	& 4063908810-  & 2.84 $\pm$ 0.41 &$ 2.66 ^{+ 0.33 }_{- 0.44 }$& 2.1-4.3 &			$3\pm0.2$, NH$^{77}$;$3.2\pm1.1$, SPH$^{61}$;				\\ &076415872&&&&$3.6$, SPH$^{65}$\\

IGR J19294+1816 	& 4323316622-  &    &$ 2.93 ^{+ 1.48 }_{- 2.47 }$& 8 &						$8$, SPH$^{68}$								\\ & 779495680 &&&& \\

KS 1947+300 		& 2031939548-  &    &$ 15.19 ^{+ 2.73 }_{- 3.67 }$& 6.0-11.3 &					$8.5\pm2.3$, SED$^{73}$;$10.4\pm0.9$, SPH$^{72}$;				\\ &802102656&&&&$8\pm2$, SPH$^{75}$;$9.5\pm1.1$, SU$^{57}$;$10$, SPH$^{51}$			\\

RX J0440.9+4431 	& 2528784015-  & 3.66 $\pm$ 0.64 &$ 3.25 ^{+ 0.45 }_{- 0.62 }$& 1.8-3.8 &			$2.9\pm0.37$, SED$^{73}$;$3.2$, SPH$^{36}$;				\\ &57369088&&&&$3.3\pm0.5$, SPH$^{56}$;$2.2\pm0.4$, SPH$^{75}$\\

SWIFT J0243.6+6124 	& 4656281935-  & 10.50 $\pm$ 3.33 &$ 6.86 ^{+ 1.11 }_{- 1.53 }$& 2.5,>5 &			$2.5\pm0.5$, SPH$^{80}$;$>5$, SU$^{79}$					\\ & 26364416 &&&& \\

SWIFT J1626.6-5156 	& 5933976985-  &    &$ 9.81 ^{+ 2.30 }_{- 3.19 }$& 7.2-15.0 &					$10.7\pm3.5$, SPH$^{70}$;$15$, SU$^{71}$				\\ & 766949376 &&&& \\

V 0332+53 		& 4447529731-  & 6.98 $\pm$ 1.74 &$ 5.13 ^{+ 0.76 }_{- 1.02 }$& 4.5-16.0 &			$6.9\pm0.71$, SED$^{73}$;$9^{+7}_{-4}$,SPH$^{26}$;			\\ &31169664&&&&$7.5\pm1.5$, SPH$^{44}$;$6\pm1.5$, SPH$^{75}$\\

Vela X-1 		& 5620657678-  & 2.61 $\pm$ 0.20 &$ 2.42 ^{+ 0.16 }_{- 0.19 }$& 1.2-2.6 &			$2.2\pm0.4$, CAII$^{4}$;$2.2\pm0.22$, SED$^{73}$;			\\ &322625920&&&&$1.3$, SPH$^{1}$;$1.2$, SPH$^{2}$;$1.9\pm0.2$, SPH$^{25}$				\\ &&&&&$2\pm0.2$, SED$^{77}$;\\

X Per 			& 1684505457-  & 0.81 $\pm$ 0.04 &$ 0.79 ^{+ 0.03 }_{- 0.04 }$& 0.4-3.7 &			$0.7\pm0.3$, ATM$^{38}$;$1.2\pm0.16$, SED$^{73}$;			\\ &92009600&&&&$0.83^{+2,87}_{-0,33}$,PX$^{40}$;$1.3\pm0.4$, SPH$^{31}$;	\\ &&&&&$0.801\pm0.138$, CAII$^{67}$;$0.8$, SPH$^{28}$;			\\ &&&&&$0.95\pm0.2$, ATM$^{43}$;\\

XTE J0658-073 		& 3052677318-  & 6.71 $\pm$ 1.98 &$ 5.11 ^{+ 0.93 }_{- 1.35 }$& 1.0-2.6 &			$1.8\pm0.8$, SPH$^{7}$							\\ &793446016&&&&\\

XTE J1946+274 		& 2028089540-  &    &$ 12.65 ^{+ 2.87 }_{- 3.89 }$& 3.2-13.7 &					$6.2\pm3$, SED$^{73}$;$12\pm1.7$, SPH$^{72}$;$9\pm1$, SPH$^{49}$;				\\ &103670144&&&&$7\pm2$, SPH$^{75}$;$9.5\pm2.9$, SU$^{52}$			\\

\label{tab:src}
\end{longtable}

\begin{table*}[h!]
\caption{Averaged relative difference of conventional distance estimation methods to GAIA distances. Values and errors correspond to the arithmetic mean and the standard derivation respectively.}
\label{tab:stat}
\centering
\begin{tabular}{ccl}
\hline \hline
Method 	&	Averaged $\frac{d_{est}-d_{BJ}}{d_{BJ}}$	&	Sample size (number of estimates)	 \\ \hline
SPH	&	+0.042	$\pm$	0.513	&	64	\\
SED	&	-0.066	$\pm$	0.350	&	17\\
ATM	&	+0.211	$\pm$	0.298	&	8\\
SU	&	+0.153	$\pm$	0.670	&	8\\
NH	&	+0.386	$\pm$	0.378	&	4\\
DH	&	-0.193	$\pm$	0.094	&	3 (Cen X-3, 4U 1538-52)\\
CAII	&	-0.039	$\pm$	0.050	&	2 (Vela X-1, X Per)\\
PX	&	-0.399	$\pm$	0.446	&	2 (1A 0535+262, X Per)\\
ACH	&	-0.685			&	1 (Cep X-4)\\
EB	&	+0.310			&	1 (Her X-1)\\
LC	&	-0.587			&	1 (2A 1822-371)	\\
RA	&	+0.086			&	1 (4U 1700-37)	\\
SV	&	-0.543			&	1 (2S 0114+650)\\
all, $d_{BJ}<5$kpc 	&	+0.182	$\pm$	0.528	&	62\\
all, $d_{BJ}>5$kpc 	&	-0.160	$\pm$	0.359	&	51\\
\hline
\end{tabular}
\end{table*}

\end{document}